\documentstyle[aps,12pt,epsf]{revtex}
\def\la{\mathrel{\mathpalette\fun <}}
\def\ga{\mathrel{\mathpalette\fun >}}
\def\fun#1#2{\lower3.6pt\vbox{\baselineskip0pt\lineskip.9pt
  \ialign{$\mathsurround=0pt#1\hfil##\hfil$\crcr#2\crcr\sim\crcr}}}

\begin{document}

\draft
\title{A Reply to \lq\lq Comment on \lq 
Big Bang Nucleosynthesis and Active-Sterile Neutrino Mixing:
Evidence for Maximal $\nu_\mu\leftrightarrow\nu_\tau$
Mixing in Super Kamiokande?\rq {\rq\rq}}

\author{Xiangdong~Shi and George~M.~Fuller}
\address{Department of Physics, University of California,
San Diego, La Jolla, California 92093-0319}

\date{November 26, 1998}

\maketitle

\begin{abstract}
In the paper \lq\lq Big Bang Nucleosynthesis and Active-Sterile
Neutrino Mixing: Evidence for Maximal Muon-Neutrino/Sterile-Neutrino
Mixing in Super Kamiokande\rq\rq\ (http://xxx.lanl.gov/abs/astro-ph/9810075),
we suggested that to evade the Big Bang Nucleosynthesis exclusion of the
muon neutrino to sterile neutrino oscillation explanation
of the Super Kamiokande data, the tau neutrino must have a
mass over about 15 eV and it must mix with a lighter sterile neutrino. 
A stable tau neutrino with this mass is inconsistent with cosmological
structure formation.  In a comment on our paper
(http://xxx.lanl.gov/abs/astro-ph/9811067),
Foot and Volkas argued that our result is incorrect and that the required
tau neutrino mass should be much lower.  Here we back up our original
result with a more detailed calculation. We show that the argument of
Foot and Volkas is invalid, most likely due to an insufficient energy
resolution in the low energy part of the neutrino spectrum.
\end{abstract}
\bigskip

\pacs{PACS numbers: 14.60.Pq; 14.60.St; 26.35.+c; 95.30.Cq}

\newpage
The Super Kamiokande atmospheric neutrino data have provided
evidence for muon neutrino oscillation as a result of a maximal
or near maximal mixing between $\nu_\mu$ and either $\nu_\tau$ or a 
sterile neutrino $\nu_s$\cite{SuperK}. The mass-squared-difference
involved in the mixing is $\sim 10^{-3}$ eV$^2$\cite{SuperK}.
The maximal or near maximal $\nu_\mu\leftrightarrow\nu_s$ mixing
explanation, however, would violate the Big Bang Nucleosynthesis
(BBN) bound by nearly completely equilibrating a fourth neutrino
flavor $\nu_s$ in the early universe and, hence, increasing the 
primordial $^4$He yield above the observed level
\cite{ShiFuller,BBNbound}.  One mechanism proposed
to evade this bound is to invoke a large lepton number asymmetry
($\ga 10^{-5}$) at the BBN epoch which acts to suppress the 
$\nu_\mu\leftrightarrow\nu_s$ transformation by matter effects.
A clever scheme along these lines suggested by Foot and Volkas
\cite{Lgenesis1} is to have a massive $\nu_\tau$ mix with a 
lighter sterile neutrino $\nu_{s^\prime}$ (which could in 
principle be the same $\nu_s$).  The resonant transformation 
of tau neutrinos to sterile neutrinos via matter-enhanced mixing
at the BBN epoch would generate a tau lepton number asymmetry
$L_{\nu_\tau}$ that grows with time \cite{Lgenesis1,Lgenesis2}.
The $\nu_\tau\leftrightarrow\nu_{s^\prime}$ resonant 
transformation must occur before any significant 
$\nu_\mu\leftrightarrow\nu_s$ transformation can occur, so that
the $L_{\nu_\tau}$ generated can subsequently suppress the
$\nu_\mu\leftrightarrow\nu_s$ transformation. However, one twist
in this scheme is that while $L_{\nu_\tau}$ grows, it crosses 
a parameter region where the MSW (Mikheyev-Smirnov-Wolfenstein)
matter effect causes either $\nu_\mu$ (if $L_{\nu_\tau}>0$) or 
$\bar\nu_\mu$ (if $L_{\nu_\tau}<0$) to resonantly transform into
$\nu_s$ or its antiparticle\cite{ShiFuller,Wrongfoot}.  The result is a
newly generated $L_{\nu_\mu}$, with a sign opposite to that of
$L_{\nu_\tau}$.  This $L_{\nu_\mu}$ acts to counter the
suppression effect of $L_{\nu_\tau}$ on the 
$\nu_\mu\leftrightarrow\nu_s$ transformation. This is because
the matter-antimatter asymmetry contribution to the effective
potential of the $\nu_\mu\leftrightarrow\nu_s$ system is 
$\propto 2L_{\nu_\mu}+L_{\nu_\tau}$ \cite{ShiFuller}.  The
countering effect of $L_{\nu_\mu}$ is immaterial only if
$L_{\nu_\tau}$ is sufficiently large at the time of the
$\nu_\mu$ or $\bar\nu_\mu$ resonant conversion. In turn, this
requires the mass-squared-difference of the 
$\nu_\tau\leftrightarrow\nu_{s^\prime}$ mixing to satisfy
$m_{\nu_\tau}^2-m^2_{\nu_{s^\prime}}\ga 300$ eV$^2$, which
implies $m_{\nu_\tau}\ga 15$ eV \cite{ShiFuller}.  Tau neutrinos
this massive are incompatible with cosmological structure
formation, and therefore would have to be unstable \cite{ShiFuller}.

In papers and in a comment to our aforementioned paper,
Foot {\sl et al.} investigated a similar problem and argued for
a much lower limit $m_{\nu_\tau}^2-m^2_{\nu_{s^\prime}}\ga 16$ 
eV$^2$\cite{Wrongfoot}. The difference between the required
threshold $\nu_\tau$ mass in our calculations and theirs may
stem from differences in the energy resolution employed. Namely,
we believe that an accurate treatment of the counter-suppression
effect of $L_{\nu_\mu}$ requires a very fine energy resolution 
in the low energy part of the mu neutrino spectrum.

We agree with Foot and Volkas that $L_{\nu_\tau}$ grows exponentially
at the initial stage of lepton number generation but then subsequently
approaches an asymptotic $T^{-4}$ growth (where $T$ is the temperature
of the universe). Foot and Volkas argued that since most of $\nu_\mu$ 
or $\bar\nu_\mu$ underwent resonances during the exponential growth
phase, there was not enough time for any significant generation of
$L_{\nu_\mu}$. Our results, on the other hand, show that the growth
rate of $L_{\nu_\tau}$ is not the central issue in the problem. 
Rather, since $L_{\nu_\tau}$ itself is small, especially during its
initial phase of exponential growth (e.g., $\la 10^{-8}$), even the
resonant conversion of a tiny fraction of $\nu_\mu$ or $\bar\nu_\mu$
into sterile neutrinos may generate a competing and significant
$L_{\nu_\mu}\approx -L_{\nu_\tau}/2$. Once such an $L_{\nu_\mu}$ has
been generated, the potential responsible for the
$\nu_\mu\leftrightarrow\nu_s$ transformation is driven close to zero,
rendering the suppression from $L_{\nu_\tau}$ ineffective. Therefore,
in order to suppress the $\nu_\mu\leftrightarrow\nu_s$ mixing required
for Super Kamiokande, the $\nu_\tau\leftrightarrow\nu_{s^\prime}$
mixing must generate an $L_{\nu_\tau}$ that is much larger than its
induced $L_{\nu_\mu}$ {\sl at any moment}.

There are two resonances involved in the problem: the
$\nu_\tau\leftrightarrow\nu_{s^\prime}$ resonance that
generates $L_{\nu_\tau}$; and the $L_{\nu_\tau}$-induced
$\nu_\mu\leftrightarrow\nu_s$ resonance that generates a
competing $L_{\nu_\mu}$. Because the effective potentials
of active-sterile neutrino mixings are neutrino energy 
dependent, at any given temperature each resonance
occurs only in a narrow energy bin in the neutrino energy 
spectra. As the temperature of the universe drops, a
particular resonance energy bin gradually moves to higher
neutrino energy, eventually sweeping across the entire 
neutrino energy spectrum. The narrow widths of the 
resonance energy bins and the rapid decrease of the 
effective mixing angles outside the resonance energy bins
enable a simple analytical/semi-analytical calculation.

In our semi-analytical numerical approach, we track the neutrino
transformations in the resonant parts of the neutrino energy
spectra. Again, these resonant regions in energy space consist
of only narrow energy bins and are functions of the temperature
and the lepton number asymmetry. This semi-analytical approach
offers two distinctive advantages: a very high energy resolution
of the neutrino spectrum (e.g.,$\Delta E/T\la 10^{-6}$); and
ease in understanding the physical processes involved,
especially the interplay between the two coupled mixing systems.
A high energy resolution is essential because the initial 
exponential growth of $L_{\nu_\tau}$ comes from minuscule 
differences between the resonance energies of the 
$\nu_\tau\leftrightarrow\nu_{s^\prime}$ system and the
$\bar\nu_\tau\leftrightarrow\bar\nu_{s^\prime}$ system, and
the tiny $L_{\nu_\tau}$ at this stage can be easily matched
by a competing $L_{\nu_\mu}$.  Energy bins that are too coarse
may therefore result in an incorrect account of the 
$L_{\nu_\tau}$ growth and an omission of the counter-balancing
effect of $L_{\nu_\mu}$ when $L_{\nu_\tau}$ is small.

Figure 1 shows our result based on the same mixing parameters
assumed in the figure of Foot and Volkas' comment
(astro-ph/9811067) to our paper. For the purpose of
comparison, no simultaneous
$\nu_\mu\leftrightarrow\nu_s$ transformation is assumed in this
figure.  Figure 1 is for the most part similar to that of
Foot and Volkas.  There are minor differences that are readily
identifiable: (1) our result tracks $T^{-4}$ more closely in the
\lq\lq power-law growth\rq\rq epoch; (2) $L_{\nu_\tau}$ in our 
results does not switch sign at the initial point of growth. The
sign difference is not surprising because of the chaotic character
of the growth, which introduces a sign ambiguity to the problem
\cite{Lgenesis2}.

As an illustration, also plotted in Figure 1, are
the $\vert L_{\nu_\tau}\vert$ required for the
$\nu_\mu\leftrightarrow\nu_s$ or $\bar\nu_\mu\leftrightarrow\bar\nu_s$
resonance to occur at $\nu_\mu$ or $\bar\nu_\mu$
energies $\epsilon_{\rm res}\equiv p^{(\mu)}_{\rm res}/T
=0.01,\,1,\,10$.  (The parameters for the
$\nu_\mu\leftrightarrow\nu_s$ mixing are
$\delta m^2=10^{-3}$ eV$^2$ and $\sin 2\theta=1$.) 
It can be seen from the figure that the lower
energy component of the $\nu_\mu$ or $\bar\nu_\mu$
neutrinos encounters the resonance first when
$\vert L_{\nu_\tau}\vert$ is very small, and
the resonance region moves through the $\nu_\mu$
or $\bar\nu_\mu$ spectrum to higher neutrino
energies as $\vert L_{\nu_\tau}\vert$ becomes larger.

For the potential proportional to $L_{\nu_\tau}+2L_{\nu_\mu}$
to successfully suppress the $\nu_\mu\leftrightarrow\nu_s$
transformation, the $L_{\nu_\mu}$ generated by the $\nu_\mu$ or 
$\bar\nu_\mu$ resonance has to be much smaller than $L_{\nu_\tau}$
in magnitude at any temperature as the $\nu_\mu$ or $\bar\nu_\mu$
resonance sweeps through the entire spectrum, i.e.,
\begin{equation}
f(\epsilon_{\rm res})\delta\epsilon_{\rm res}\,R\,
\left\vert{\delta\epsilon_{\rm res}
\over\dot\epsilon_{\rm res}}\right\vert
< {4\over 3}\left(L_{\nu_\tau}+2L_{\nu_\mu}\right)
\label{requirement}
\end{equation}
for any $\epsilon_{\rm res}$.
In the equation, $f$ is the Fermi-Dirac distribution function.
$\delta\epsilon_{\rm res}$ is the energy width of the resonance.
$f(\epsilon_{\rm res})\delta\epsilon_{\rm res}$
is therefore the fraction of mu neutrinos in resonance. $R$ is
the resonant transition rate.
$\vert\delta\epsilon_{\rm res}/\dot\epsilon_{\rm res}\vert$
is the duration of the resonance at $\epsilon_{\rm res}$.
The energy width of the resonance
depends on whether the resonant transition is collision-dominated
(with the quantum damping rate $D\sim 0.5G_F^2T^5\epsilon_{\rm res}>V_x$)
or oscillation-dominated ($D<V_x$):
\begin{equation}
\delta\epsilon_{\rm res} \sim \left\{  \begin{array}{ll}
2\left\vert D{\partial\epsilon_{\rm res}/\partial V_z}\right\vert &
\mbox{if $D>V_x$}\\
2\left\vert V_x{\partial\epsilon_{\rm res}/\partial V_z}\right\vert &
\mbox{if $D<V_x$}\end{array}\right..
\label{width}
\end{equation}
So does the resonant transition rate:
\begin{equation}
R\approx \left\{  \begin{array}{ll}
V_x^2/D & \mbox{if $D>V_x$}\\
V_x     & \mbox{if $D<V_x$}\end{array}
\right..
\label{transitionrate}
\end{equation}

The effective potentials for the
$\nu_\mu\leftrightarrow\nu_s$ transformation are\cite{ShiFuller}
\begin{equation}
V_x=
{\vert m_{\nu_\mu}^2-m_{\nu_s}^2\vert\over 2\epsilon T}\sin 2\theta,
\quad V_z\approx 22G_F^2T^5\epsilon\pm 0.35G_FT^3(L_{\nu_\tau}+2L_{\nu_\mu}),
\label{potential}
\end{equation}
where $G_F$ is the Fermi constant, the
\lq\lq $+$\rq\rq\ (\lq\lq $-$\rq\rq ) sign is for $\nu_\mu$
($\bar\nu_\mu$), and we employ natural units. (Here $V_y=0$.)
The $\nu_\mu$ or $\bar\nu_\mu$ resonance occurs at an energy
$\epsilon_{\rm res}\approx (L_{\nu_\tau}+2L_{\nu_\mu})/63G_FT^2$, with
$\partial\epsilon_{\rm res}/\partial V_z\approx (22G_F^2T^5)^{-1}$.
The temperature $T_{\rm res}$ at which the resonance occurs
is almost independent of $\epsilon_{\rm res}$:
$T_{\rm res}\approx 22[(m_{\nu_\tau}^2-m_{\nu_{s^\prime}}^2)/
1\,{\rm eV}^2]$. ($T_{\rm res}$ is an insensitive function of
the $\nu_\tau\leftrightarrow\nu_{s^\prime}$ vacuum mixing angle.)
Furthermore, we only consider the case where $\epsilon_{\rm res}\ll 1$.
This is when $L_{\nu_\tau}$ is in its
initial stage of growth ($L_{\nu_\tau}\ll 10^{-7}$) and is
most easily matched by a competing $L_{\nu_\mu}$.
Given the small $\epsilon_{\rm res}$
we have $f(\epsilon_{\rm res})\approx \epsilon_{\rm res}^2/1.8$.
Eq.~({\ref{requirement}) can then be rewritten in the form
\begin{equation}
{\vert m_{\nu_\mu}^2-m_{\nu_s}^2\vert^2
\over 1600G_F^2T_{\rm res}^7\vert\dot\epsilon_{\rm res}\vert}
<80G_FT_{\rm res}^2.
\label{eq3+}
\end{equation}
if $D>V_x$, or
\begin{equation}
{\vert m_{\nu_\mu}^2-m_{\nu_s}^2\vert^3\over 
1800G_F^4T_{\rm res}^{13}\epsilon_{\rm res}^2\vert\dot\epsilon_{\rm res}\vert}
<80G_FT_{\rm res}^2.
\label{eq3-}
\end{equation}
if $D<V_x$.

We can further rewrite $\vert\dot\epsilon_{\rm res}\vert\equiv 
H\epsilon_{\rm res}\vert {\rm d}\ln\epsilon_{\rm res}/{\rm d}\ln T\vert$
where $H=-{\rm d}\ln T/{\rm d}t\approx 5.5T^2/m_{pl}$ is the Hubble 
expansion rate. The Planck mass is $m_{pl}\approx
1.22\times 10^{28}$ eV. Then Eq.~(\ref{eq3+}) becomes
\begin{equation}
\left({m_{\nu_\tau}^2-m_{\nu_{s^\prime}}^2\over 1\,{\rm eV}^2}\right)^{11/6}
>2\times 10^4\epsilon_{\rm res}^{-1}
\left\vert {{\rm d}\ln\epsilon_{\rm res}\over {\rm d}\ln T}\right\vert^{-1}
\left\vert {m_{\nu_\mu}^2-m_{\nu_s}^2\over 10^{-3}\,{\rm eV}^2}\right\vert^2 .
\label{main+}
\end{equation}
for $D>V_x$.  And  Eq.~(\ref{eq3-}) becomes
\begin{equation}
\left({m_{\nu_\tau}^2-m_{\nu_{s^\prime}}^2\over 1\,{\rm eV}^2}\right)^{17/6}
>10^3\epsilon_{\rm res}^{-3}
\left\vert {{\rm d}\ln\epsilon_{\rm res}\over {\rm d}\ln T}\right\vert^{-1}
\left\vert {m_{\nu_\mu}^2-m_{\nu_s}^2\over 10^{-3}\,{\rm eV}^2}\right\vert^3 .
\label{main-}
\end{equation}
for $D<V_x$.

The value of $\vert {\rm d}\ln\epsilon_{\rm res}/{\rm d}\ln T\vert$
is related to the growth of $L_{\nu_\tau}$
by $\vert {\rm d}\ln\epsilon_{\rm res}/{\rm d}\ln T\vert
\approx\vert {\rm d}\ln L_{\nu_\tau}/{\rm d}\ln T -2\vert
\approx\vert {\rm d}\ln L_{\nu_\tau}/{\rm d}\ln T\vert$ 
(with $L_{\nu_\mu}$ safely ignored).  Figure 2 shows
$\vert {\rm d}\ln L_{\nu_\tau}/{\rm d}\ln T \vert$
as a function of the $\nu_\tau\leftrightarrow\nu_{s^\prime}$
vacuum mixing parameters, $m^2_{\nu_\tau}-m^2_{\nu_s}$
and $\sin^22\theta^\prime$, in the initial exponential stage of
$L_{\nu_\tau}$ growth when $L_{\nu_\tau}$ is $\ll 10^{-7}$.
$\vert {\rm d}\ln L_{\nu_\tau}/{\rm d}\ln T \vert$
can be approximately fit as
\begin{equation}
\left\vert {{\rm d}\ln L_{\nu_\tau}\over {\rm d}\ln T }\right\vert
\approx 6\times 10^6\sin2\theta^\prime,
\label{upperbound}
\end{equation}
and is insensitive to $m^2_{\nu_\tau}-m^2_{\nu_s}$.
The vacuum mixing parameter $\sin2\theta^\prime$ of the
$\nu_\tau\leftrightarrow \nu_{s^\prime}$ mixing must satisfy
the BBN bound $(m_{\nu_\tau}^2-m_{\nu_{s^\prime}}^2)\sin^42\theta^\prime
<10^{-7}\,{\rm eV}^2$ (the cut-off at the upper-right corner of Figure 2)
\cite{ShiFuller}.

Eq.~(\ref{main+}),~(\ref{main-}) and (\ref{upperbound}) show that
the most stringent requirement on $m_{\nu_\tau}^2-m_{\nu_{s^\prime}}^2$
does indeed come not from $\nu_\mu\leftrightarrow\nu_s$ resonances
at $\epsilon_{\rm res}\sim 3$, but from resonances centered
at the smallest possible $\epsilon_{\rm res}$ as long as the
$\nu_\mu$ or $\bar\nu_\mu$
transition probability in that resonance energy bin is $\ll 1$.
This condition, expressed as
\begin{equation}
R\left\vert{\delta\epsilon_{\rm res}\over\dot
\epsilon_{\rm res}}\right\vert\la 0.1,
\end{equation}
can be rewritten as
\begin{equation}
\epsilon_{\rm res}\ga 
\left({m_{\nu_\tau}^2-m_{\nu_{s^\prime}}^2\over 1\,{\rm eV}^2}\right)^{-1/2}
\left\vert {{\rm d}\ln\epsilon_{\rm res}\over {\rm d}\ln T}\right\vert^{-1/3}
\left\vert {m_{\nu_\mu}^2-m_{\nu_s}^2\over 10^{-3}\,{\rm eV}^2}
\right\vert^{2/3},
\label{howsmall}
\end{equation}
regardless of the value of $D/V_x$.
It can be shown that $\epsilon_{\rm res}$ is
in the oscillation-dominated regime if
\begin{equation}
\epsilon_{\rm res}\la 0.25 \left\vert
{m_{\nu_\mu}^2-m_{\nu_s}^2\over 10^{-3}\,{\rm eV}^2}\right\vert^{1/2}
\left({m_{\nu_\tau}^2-m_{\nu_{s^\prime}}^2\over 1\,{\rm eV}^2}\right)^{-1/2}.
\end{equation}
Therefore, the most stringent requirement on
$m_{\nu_\tau}^2-m_{\nu_{s^\prime}}^2$ comes from the
oscillation-dominated regime for $\sin^22\theta^\prime\ga 10^{-8}$ 
(while $\vert {{\rm d}\ln\epsilon_{\rm res}/{\rm d}\ln T}\vert\ga 10^3$),
and from collision-dominated regime for $\sin^22\theta^\prime\la 10^{-8}$ 
(while $\vert {{\rm d}\ln\epsilon_{\rm res}/{\rm d}\ln T}\vert\la 10^3$).

Combining Eq.~(\ref{main+}) or~(\ref{main-}), Eq.~(\ref{upperbound})
and Eq.~(\ref{howsmall}) yields a requirement on the 
mass-squared-difference necessary to effect suppression
of $\nu_\mu\leftrightarrow\nu_s$ transformation at the
Super Kamiokande level:
\begin{equation}
m_{\nu_\tau}^2-m_{\nu_{s^\prime}}^2\ga \left\{ \begin{array}{ll}
200\left(\vert m_{\nu_\mu}^2-m_{\nu_s}^2\vert /10^{-3}\,{\rm eV}^2\right)
^{3/4}\,{\rm eV}^2 & \mbox{for $\sin^22\theta^\prime\ga 10^{-8}$}\\
\left(\sin 2\theta^\prime\right)^{-1/2}
\left(\vert m_{\nu_\mu}^2-m_{\nu_s}^2\vert /10^{-3}\,{\rm eV}^2\right)
\,{\rm eV}^2 & \mbox{for $\sin^22\theta^\prime\la 10^{-8}$}\end{array}
\right.
\label{master}
\end{equation}
This is in agreement with our previous work.

Perhaps because of insufficient energy resolution
at low energies, Foot and Volkas may have missed
the crucial effect of small $\epsilon_{\rm res}$.
Neutrinos with energies $\sim 1$\% of the 
temperature are an insignificant fraction ($\sim 10^{-6}$)
of the overall neutrino number.  However, 
in this problem, they are the driving force
which frees the $\nu_\mu\leftrightarrow\nu_s$
transformation process.  This is simply because
the suppressing lepton asymmetry from $\nu_\tau$
in this case is itself minuscule (e.g.,$\ll 10^{-7}$).

\bigskip
\noindent{\bf Figure Captions:}

\noindent
Figure 1. The growth of the tau neutrino asymmetry as
a result of the tau neutrino-sterile neutrino mixing, assuming
$m^2_{\nu_\tau}-m^2_{\nu_{s^\prime}}=50$ eV$^2$ and
$\sin^2 2\theta^\prime=10^{-4}$.
The intersections between the growth curve of $L_{\nu_\tau}$ and
the dashed lines indicate when resonances occur for
$\nu_\mu$ (if $L_{\nu_\tau}>0$) or $\bar\nu_\mu$ (if $L_{\nu_\tau}<0$)
neutrinos with momentum $p$.
\bigskip

\noindent
Figure 2. The initial rate of $L_{\nu_\tau}$ growth, 
${\rm d}\ln L/{\rm d}\ln T$, as a function of the 
vacuum tau neutrino-sterile neutrino mixing parameters.
\bigskip

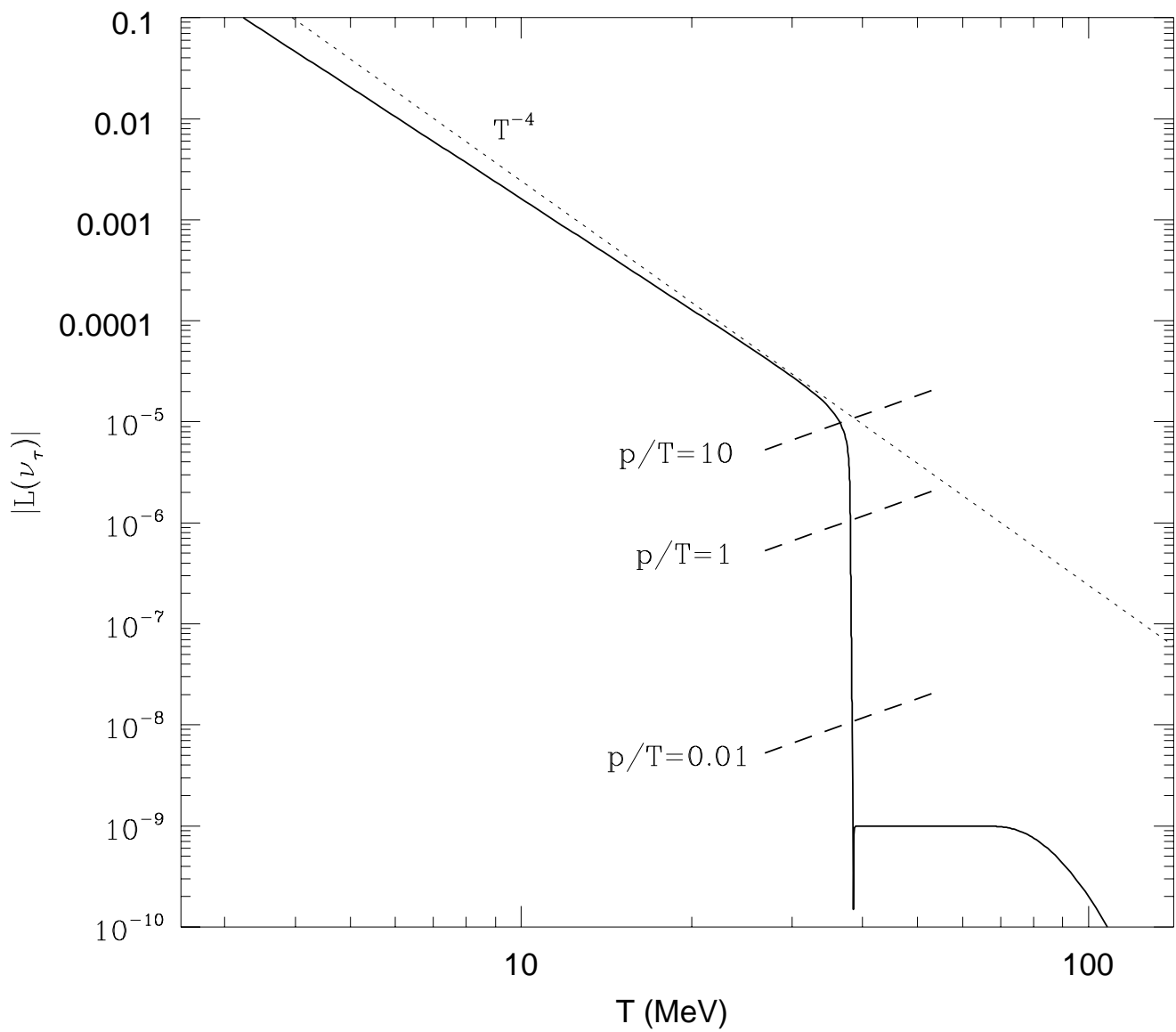
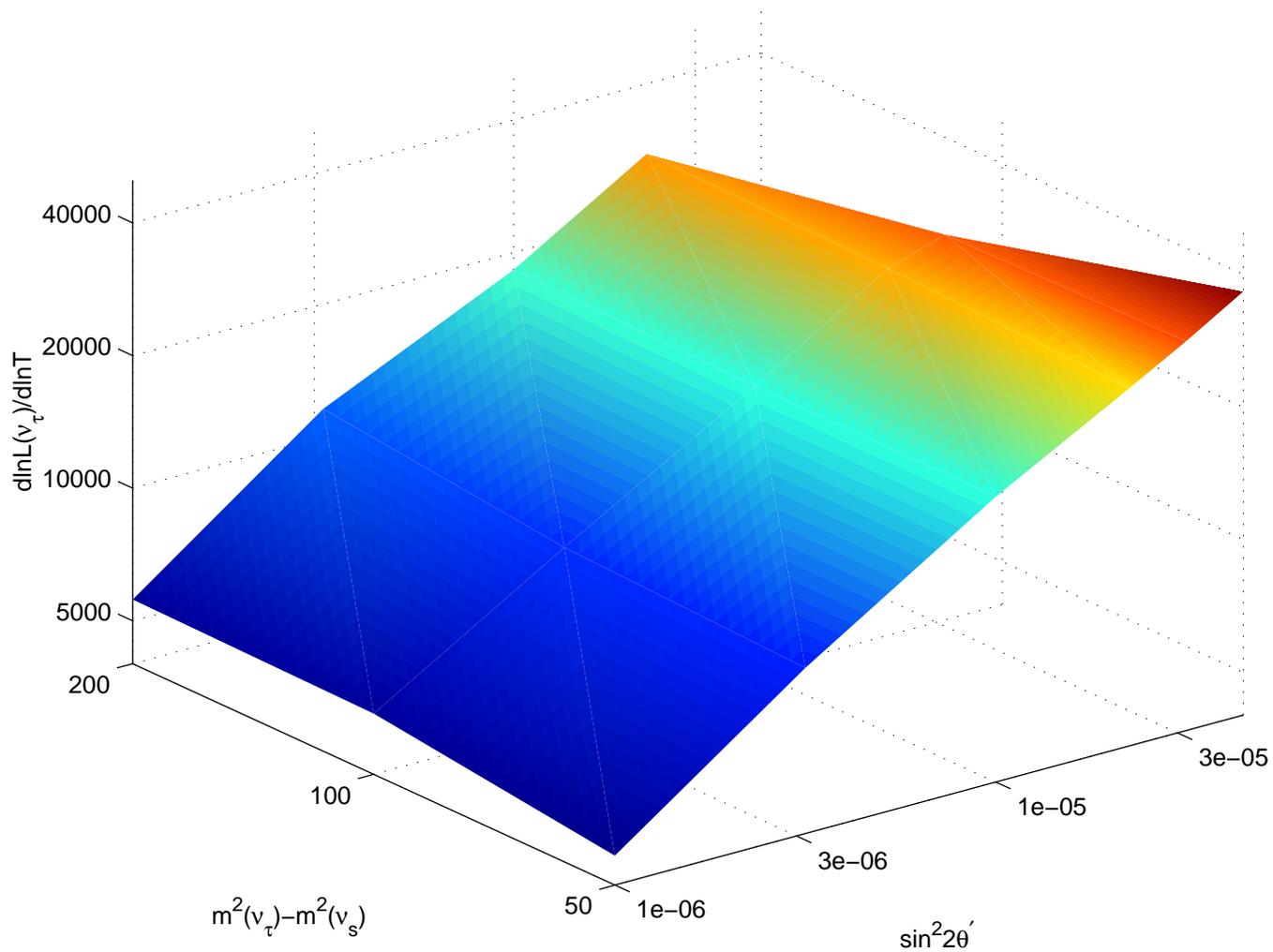
\end{document}